\documentclass[lnicst, 10pt]{svmultln}

\usepackage{makeidx}
\usepackage{verbatim}
\usepackage{amsmath}
\usepackage{amssymb}
\usepackage{bbm}
\usepackage{graphicx}
\usepackage{url}

\usepackage{amssymb}
\usepackage{graphicx}
\usepackage{subfigure}
\usepackage{color}
\usepackage{comment}
\usepackage{amssymb,amsmath,latexsym,amsfonts}

\usepackage{url}
\urldef{\mailsa}\path|{eitan.altman,mhanawal}@inria.fr, yuedong.xu@gmail.com, jrojasmo7@alumnes.ub.edu,swong@udc.es|

\begin{document}

\mainmatter  

\title{Net Neutrality and Quality of Service}

\titlerunning{Net Neutrality and Quality of Service}

%
%
\author{Eitan Altman$^1$%
\and Julio Rojas$^{1,2}$ \and Sulan Wong$^{1,3}$
\and Manjesh Kumar Hanawal$^{1,4}$ \and Yuedong Xu$^4$}
\authorrunning{E. Altman et al.}

\institute{$^1$ INRIA Sophia Antipolis, 2004 Route des Lucioles, France \\
$^2$Dept. of Econ. and Bus. Sci. at Univ. of Barcelona,
08034 Barcelona, Spain   \\
$^3$Dept. of Law, Univ. of Coru\~na, 15071 A Coru\~na, Spain
\\
$^4$LIA, University of Avignon, 339, chemin des Meinajaries, Avignon, France
 \mailsa}


%
%


\maketitle

\begin{abstract}
{\small
2010 has witnessed many public consultations around the world
concerning Net neutrality. A second legislative phase
that may follow, could involve various structural changes in the
Internet. The status that the Internet access has in Europe as a universal
service evolves as the level of quality of service (QoS) to be offered
improves. If guarantees on QoS are to be imposed,
as requested by several economic actors,
it would require introducing new indicators of quality of services,
as well as regulation legislation and monitoring  of the
offered levels of QoS. This tendency in Europe may change
the nature of the Internet from a best effort network to, perhaps, a
more expensive one, that offers guaranteed performance.
This paper presents an overview of the above issues as well
as an overview of recent research on net-neutrality, with an
emphasis on game theoretical approaches.
}
\end{abstract}

\section{Introduction}
Several public consultations on network neutrality policies
have taken place in 2010. 
From them, regulation was introduced by the FCC in the USA \cite{FCC:2010c}, while
the European authorities feel there is no need for it \cite[p. 3]{EuroConsultNNRep2010} at this moment.
France's Assembly, meanwhile, is discussing a bill on net neutrality \cite{France3061}.
The growing economic and social role
of the Internet along with the fast evolution of its performance
and of the services it offers, have triggered evolution of the legal
status of the access to the Internet. Already on 2002,
access to the Internet has become an universal service in the
EU, which should imply guarantees on QoS. Even if
data connections ``should be capable of supporting data communications
at rates sufficient for access to online services such as those provided
via the public Internet'', the European Parliament understands that due
to the heterogeneous nature of the Internet ``it is not appropriate to
mandate a specific data or bit rate at Community level'', leaving Member States
with the responsibility of monitoring if the data rates provided by ISPs are ``sufficient
to permit functional Internet access'' \cite[Whereas 5]{Directiva136:2009a}.
Nonetheless, this guarantee only covers access to the narrowband Internet service, while
the definition of the broadband Internet service as a universal service
is currently in discussion \cite{Kroes:2010b}.

From the debate
on network neutrality we learn that new indicators of QoS are sought, and that we should expect an intensive work
of regulating and standardization bodies on
defining requested minimum values of performance measures.
By actually requiring QoS to be guaranteed
to the end user's actual experience (see e.g.  \cite{arcep-may2010e}
page 19), we may expect new legislation to create a  new
reliable and yet more expensive Internet, that would be different
than the best effort type network that we have known.

This issue of a best effort versus a guaranteed performance
network was not only present in the debate on
network neutrality, but could be central in the legislation
that would follow that debate. Another new comer issue is that
of imposing taxes on content providers by the government.
We provide more details on both issues in the next section.

The paper is organized as follows.
The next section provides an overview on (i) the
Internet as a universal service, (ii) the legislation concerning
QoS, and on  (iii)
economic issues that arise in the debate on network neutrality.
The following section provides a brief non exhaustive overview
on mathematical models related to the neutral net question, with
special focus on game theoretical models.

\section{Overview}

{\bf The economic aspects.}
The information and communication technologies have a central
economic role. Indeed,
"The information and communication technologies sector already generates revenue
of 2,700 billion euros, or close to 7\% of global GDP, and could account
for 20\% of GDP within the next 10 years" \cite{arcep-may2010e}.
The latter reference specifies moreover:
"Pyramid Research and Light Reading
predict a rise in
annual worldwide revenue for voice and data services of around 2.5\%
and 12.8\%, respectively, between 2010 and 2013, while data traffic is
forecast to increase by
131\% during that same period. To give an example, in December 2009 ComScore
reported that 5.4 billion videos had been watched in France that year (a 141\%
increase over the year before), of which 1.8 billion on YouTube.com between
January and September".

In Europe, 5\% of its GDP, i.e.
 660 billion annually, comes from the IT sector, with
250 million daily Internet users and a penetration of the mobile market close to 100\% \cite{COM_2010_245}.

{\bf Internet as a tool for exercising the freedom of speech.}
In 2009, France passed a law against non-authorized downloading
of copyrighted material. Measures against file-sharing included
disconnection from the Internet through an administrative order.
The Constitutional Council went back to the
{\bf Declaration of the Rights of Man and of the Citizen}
(from the time of the French revolution, two hundred years
before the Internet was born) to conclude that freedom
of speech could not be trusted to a new nonjudicial
authority in order to protect holders of copyrights and neighboring rights.
In their judgement, it recognized that the Internet is an instrument
for exercising the freedom of speech. Similar relations between the
Internet and the American constitution (and amendments) have also
been made in the USA, see \cite{comnet} for more details.

Recent events in the Arab world have shown us that people use
Internet as a source not only to express their opinion on governments,
but also to coordinate actions that allows overthrowing them.
Governments have reacted, predictably, by blocking not only the Internet
but also the mobile phone service. However, the effectiveness of these
measures has been compromised with the release in media outlets
throughout the world, of videos recorded in those countries.

{\bf Internet Access as a universal service.}
Directive 2002/22/EC of the European Union established the access
to the Internet as a universal service, i.e. a global service guaranteed
to all end users, regardless of their geographical location, at reasonable
quality and reliability and at affordable price.
Directive 2009/136/CD did not only request that access
has a reasonable guaranteed service quality, it further
complemented this request by giving ANRs the power to define
the minimum QoS levels, in order to
avoid service degradation, throttling and blocking.



To apply the law, one needs to constantly
monitor the QoS in order to check if the targets
values are indeed met. It is our conviction that this monitoring
should involve not only those who offer the services and access to
the Internet but also scientists that come from the public sector
(research institutes and universities) who do not have direct
economic interests.

\subsection{New indicators of quality of service needed}
\label{sec:q1}

Since 2002, there has been an exponential increase in
the internet traffic, with new services and applications that
appeared.  In view of this huge growth, it has become clear that levels of
QoS that have been appropriate for 2002 are  no more
sufficient, and moreover, there is a need in new indicators of QoS.

On the legislation side, there is a need to redefine the quality
of services which the European universal service should offer.
This can be done also in legislation at a state level.
The first such initiative has been taken by Finland where
already on October 2009 \cite{MinistryFinland:2009},
followed by Spain in February 2010 \cite[Art. 53]{EconSost2011}.
Both of them require from providers a minimum guaranteed rate
of 1 Mbps, a rate that makes difficult to see how the new multimedia services
that are available through Internet could be provided.

Not surprisingly, we find the issue of QoS of today's
and of future Internet in the center of the debate on network neutrality.

In \cite{arcep-may2010e},
ARCEP (the French regulation body of Electronic Communications Markets)
proposes six general policy directions for the network neutrality policy.
The third one concerns QoS and it is summarized by:
"3rd direction: A connection to the Internet must be provided with
a sufficient and transparent QoS.  To guarantee this, the
Authority is launching sector-specific efforts to qualify the minimum
QoS parameters for Internet access, and is working
to implement specific indicators."

The document further specifies: "End users must be
contractually informed of the technical properties of their Internet
access, so that they can know the resources that have been assigned
to them and the performance they can expect under "normal conditions"
(i.e. "best effort" operations)...
Work also needs to be done on the contribution of other players in the equation
(ISPs\footnote{Internet Service Providers},
equipment manufacturers, software providers, etc.)".

How should one proceed to determine the required indicators
of QoS as well as their minimum value?

In the paragraph on QoS (related to the third proposed
policy direction in \cite{arcep-may2010e}), ARCEP first invites
"operators and the associations that represent them to
engage in sector-specific work devoted to setting minimum QoS
parameters for "Internet access" (availability, bandwidth,
latency, packet loss, jitter, etc.)."

They then propose that "this work could be the basis of exchanges
with consumer associations and be enhanced by close collaboration
with other relevant players, and particularly
with ISPs since, as
the designers of services and applications, they
are particularly well suited to analyze user's qualitative experience."

Would it indeed be enough to put together the operators  with
the consumer associations in order to come up with indicators and
minimum levels of QoS?

Do the consumer associations have the experts to
understand the impact of choices of minimum values of quality
of services, and of proposed indicators, on
the quality they would perceive? Do the legislators have these?

We recommend to involve in this work a third actor, such as
research institutes and universities etc.,
that has
the required experts in answering these questions and has the commitment
of contributing to progress of the society.

\subsection{The involvement of end-users in determining policies in France}

2010 has experienced dramatic events related to network neutrality
question. Exceptional legislation initiatives have been taken,
that may pave the way to shape a different future Internet.
2010 saw the first country, Chile, adopting a legislation
that establishes network neutrality. At the same year, a
USA Court shook the foundation of the Net
neutrality in USA by denying the USA telecom regulation body,
the  FCC, the authority to take decisions and actions on that topic.
In preparation to legislation on the topic, public consultations were
launched in USA, France and the European Union (EU).

There is a huge difference in the number of participants between the
American consultation, on one side, and the French and European ones,
on the other. In total there are more than 89,000 filings in the American
one, where as the French consultation was answered by only 121 stake holders
\cite{FrenchConsRes2010} and the European one was answered 318 times
\cite{EuroConsultNNRep2010}. As can be seen in the FCC web site,
the vast majority of the answers to the consultation in USA came from
individuals who used a web tool provided by the platform \url{savetheinternet.com}
for the automation of this process. This tool had a very basic template
with a short standard text in favor of net neutrality, in which the
interested individual provided his name.

If the level of participation seems low for France, a country with
close to 65 million people of which 68.9 \% have access to Internet
\cite{InternetEU2010}, we were shocked by the number of responses
that the European one attracted, as the EU has an estimated two thirds
more population than the USA.

Among the 121 responses made by stake holders \cite{FrenchConsRes2010}
in the French consultation, eight came from ISPs, four from networking
vendors, six from content production corporations, three from copyright
collecting societies, eleven from software and content providers,
six from user associations, three from public initiative networks,
four from other kinds of professional associations, two from a group
of experts gathered by Nathalie Kosciusko-Morizet%
\footnote{French State Secretary for the Digital Economy.%
}, five from researchers, and 67 from individual citizens.

The European call also showed low level of interest among stake holders,
attracting a total of 318 responses \cite{EuroConsultNNRep2010}.
Of this relatively small participation, 34 answers came from ISPs,
seven from infrastructure providers, two from mobile phone manufacturers,
16 from national and regional authorities, 38 from Internet-related
organizations, 18 from media-related organizations, six from content
providers, 42 from industrial organizations, social, consumer and
nongovernmental organizations, seven from other companies, ten from
academic institutions, three from political parties, and 145 from
individual citizens. 

We can compare these figures with the consultation process in Canada
that was initiated by the CRCT. It provided around three months for
sending comments to the commission. In addition it had several days
of hearing. The commission which received {}``437 initial comments,
35 reply comments, and 34 final replies from parties (companies and
advocacy groups) and individuals. In addition, an online campaign
resulted in over 13,000 email submissions to the Commission from individuals.
At the oral hearing in July 2009, 26 presentations were made. Finally,
an online consultation initiated by the Commission resulted in 1,400
additional individual comments'' \cite[\S 10]{Canada_Consultation:2009}.
Thus the response to both, the French and the European consultations,
are also much lower than the one for the Canadian consultation.

\subsection{Duration and timing}

Of the three public consultation processes, the French was the shortest,
running for 39 days (April 9 to May 17, 2010), followed by the European,
that extended for 93 days (June 30, 2010 to September 30, 2010), and
then the longest is the American, which ran for 187 days (October
22, 2009 to April 26, 2010)%
\footnote{Due to the {}``Comcast'' decision\cite{COMCAST_v_FCC:2010}, the
FCC extended the deadline for filing reply comments in response to
the NPRM from March 5 to April 26.%
}. The French consultation is not only much shorter than the American.
The timing for the consultation was chosen to overlap the two weeks
vacation period of the Eastern holidays in France, in which many French
spending vacations with their families are disconnected from the politics.
Since both the duration and the timing of the consultation are under
the control of the government, it seems natural to speculate that
the French government was not interested in having a large participation.
The European consultation was not only half as long as the American
one, it also had the same problem of the French one, as it ran over
the summer holidays that usually spread from mid-July to mid-September,
a period of almost 60 days of very little activity by the consultees.

\subsection{The impact of the government position}

As both, the U.S. government led by President Obama and the FCC headed
by the Commissioner Genachoswki, have been strong proponents of the
consecration of the principle of net neutrality, either by an amendment
of the Telecommunications Act, or by an administrative mandate issued
by the FCC itself, the debate on net neutrality has been re-launched,
achieving a media presence that is usually very difficult to reach
for such a complex issue that weaves together three different areas
of knowledge. In December 2010, the FCC issued a Report and Order \cite{FCC:2010c}
as the conclusive and regulatory document obtained from the NPRM.
In it, the FCC keeps maintaining its authority to adopt rules on the
open Internet \cite[Part IV]{FCC:2010c}, but not unanimously, as
two of its members believe that it does not so \cite[p. 148-150, 188-193]{FCC:2010c}.

The European Union, both through the statement issued by the Commission
under the Telecom Package%
\footnote{The Commission acknowledges in a declaration attached to the Telecom
Package that net neutrality is {}``a policy objective and regulatory
principle to be promoted by national regulatory authorities'' \cite{Commission-of-the-European-Communities:2009}.%
} as well as through Commissioners Reding and Kroes, made it clear
that the intention of the European government is to protect the neutrality
of the network.

We recall that in France, the conditions chosen to launch the consultation
seem to indicate that the government was not interested in receiving
a large number of responses. Minimizing the dimensions of the public
debate associated with the consultation may also be useful in order
to avoid the French citizens questioning other aspects of the government
policy on the Internet. In particular, the government was probably
aware that public discussions on the HADOPI law%
\footnote{Adopted in France last year and which bans downloading unauthorized
copyrighted content.%
} could be triggered by the fact that the questionnaire of the French
consultation includes an important link between the HADOPI law and
the proposed net neutrality. The relatively limited interest in France
in getting feedback from end users should not be interpreted as a
disinterest in the opinion of the various economic actors. Indeed,
as we have already seen, there was a conference held by the ARCEP
in which important economic actors participated%
\footnote{We were surprised not to see among the participants speakers from
French universities or research institutes. In fact, the only two
talks from Professors from universities are from the USA. In addition,
one can find video interviews of many stake holders in the conference's
home page: \url{http://www.arcep.fr/index.php?id=10370}. %
}.

\subsection{Taxation issues}

Among the issues that Net neutrality is concerned with are
relations between access and content providers along with
related pricing issues, as well as the possibility of an access provider to
have exclusive agreement with some content provider or some service
provider.

In Europe, these issues take another dimension, due to the fact that
many large content  providers (such as google, facebook etc)
are non-european: these are mainly American companies.
These companies make large benefits from advertisement.

It is known that google pays very
little taxes on this income, in contrast to what google pays in the USA.
In a context where all other actors related to the Internet
do pay taxes, applying network neutrality would mean favoring
or subsidising these providers that do not pay.
To be more precise, Google managed to cut 3.1 billion American
dollars of taxes in the three last years by declaring its foreign
profits (made in Europe) in Bermuda. This enabled Google to reduce its
overseas tax rate to 2.4\% . This is done using Irish law that allows to
legally shuttle profits into and out of subsidiaries there, thus escaping
the Irish 12.5 \%  tax. Facebook is preparing a similar strategy to
shift declared benefits from Ireland to Caymans. For details, see
\cite{taxgoogle}.

	The so-called {}``Zelnick Report'' \cite{ZelnikToubonCerutti2010},
which came out in France in January 2010, proposed to impose a
tax on advertising
revenue generated by the use of online services from France. According
to estimates put forward by the authors, between 10 to 20 million
euros would be collected mainly from U.S. content providers (Google,
Microsoft, AOL, Yahoo and Facebook). It is pertinent to note that
the report expresses concerns about the drop in advertising revenues
of the French content providers, citing the poor state of competition
in the French market for search engines, and certain behaviors (never
clarified in the text) of Google. Later, French President Nicolas
Sarkozy supported this proposal in a speech where he presented a set
of policies to support the sector of cultural content
creation.

The French position concerning google had perhaps some impact.
In September 2010, Google CEO Eric Schmidt met with the French president
Sarkozy to discuss
the opening in Paris of a Google research center and the creation
of a European cultural institute\cite{Google2010}. Schmidt said that
the only reason for this initiative is economic, since it considers
the French market for online searches as very dynamic. He added that
in the meeting, the so-called {}``Google tax'' was not discussed
and that he has not met with the French competition authority.

\subsection{Incentives for Investments}

One of the issues in the debate on Network Neutrality
has been incentives for investments in the infrastructure. Some argue
that neutrality would create the incentives and some argue that
only a non--neutral net would guarantee that.
This problem has been partly resolved in the European Union when Internet
access was declared as a universal service. Indeed, there are
several possible ways to finance the cost of
Providing communications services to all end-users comes.
The Universal Service Directive allows providers to be compensated
either from public funds or through a cost-sharing arrangement between
providers if it is demonstrated that by complying with the
universal service obligations they incur a loss or suffer
net costs that exceed normal commercial standards \cite{universal}.

The latter reference further says:
"Member States are free to go beyond the minimum requirements
laid down in the Directive, the only stipulation being that
any additional obligation cannot be funded by a Levy on
telecom providers."

We note however that in order to guarantee that the
development of infrastructure would continue, it may be necessary
to upgrade also the legislation on the universal service in order
to apply it not just for the Internet access but also to
broadband Internet services, which has not yet been
declared universal service by the European community.

\section{Mathematical modeling of Network Neutrality Issues}

\subsection{Conclusions of the models at a Glance}

There is one particular economic issue that is at the heart of the
conflict over network neutrality. Hahn and Wallsten
\cite{HW} write that net neutrality ``usually means that broadband
service providers charge consumers only once for Internet access, do
not favor one content provider over another, and do not charge
content providers for sending information over broadband lines to end
users.'' This motivates the recent studies on the implications
of being non-neutral and of charging the content providers (CP).
%

Two central questions in the context of a possibly
non-neutral Internet which many researchers have tried to
to answer are \cite{CBG}: i) who would gain  or loser in the absence of neutrality;
ii) will ISPs or CPs have more incentives to invest on
the network infrastructure.
The heated debate on these questions involve legal, economic
and technological aspects, see e.g.  \cite{economides}.
There have not been much work which involves all these expertise.
Moreover, some  of the existing works draw contradictory conclusions
due to their differences in the market modeling.

%
%

We first take a glance at the state of the art research before diving
into the details.
We concentrate on the profits of the economic actors as well as the incentive of
investment when the network neutrality is abandoned. The main
results of existing work are
summarized in Table \ref{table:comp}. The first column lists the recent
work as well as the year of publication. The second column
highlights the power of ISPs in an Internet market.
The charging of CPs may have different impacts on network utility in
the competing and the monopoly ISP markets. The symbol $\surd$ denotes that the
performance metric is better off with network non-neutrality.
On the contrary, the symbol $\times$ represents a disadvantage in the non-neutral networks.
For the two questions raised above, only part of existing works present definite answers.
We use the symbol $\star$ to denote the case where the authors provide
a more complex result: they  provide some conditions under which
the network non-neutrality
is beneficial and others for which it is harmful. The symbol $\varnothing$
means that the authors have not studied that specific metric.
From this table, we can conclude that the available models do
not seem to agree with each other. However, the ISPs have more incentives
to invest if they are better off with no regulation, according to these studies (including
\cite{networking2011,Walrand09}).


\begin{table}[!htb]
\centering
\begin{tabular}[width=0.9\linewidth]{|c|c|c|c|c|c|c|}
\hline \hline  & Market & ISP & CP & User & Social & ISP\\
&  & Surplus & Surplus & Surplus & Surplus & Investment\\
\hline ET\cite{Economides09} & One ISP & $\surd$ & $\times$ & $\surd$ & $\times$ & $\varnothing$\\
\hline ET\cite{Economides09} & Two ISPs & $\times$ & $\times$ & $\surd$ & $\times$ & $\varnothing$\\
\hline NOSW\cite{asu} & Two ISPs & $\surd$ & $\surd$ & $\surd$ & $\surd$ & $\surd$\\
\hline CBG\cite{CBG} & One ISP & $\surd$ & $\times$ & $\surd$ & $\surd$ & $\times$\\
\hline JL\cite{JL} & One ISP & $\varnothing$ & $\varnothing$ & $\varnothing$ & $\surd$ & $\surd$\\
\hline MSW\cite{Walrand09} & More ISPs & $\star$ & $\star$ & $\star$ & $\star$ & $\star$ \\
\hline ALX\cite{networking2011} & One ISP & $\star$ & $\star$ & $\star$ & $\star$ & $\star$ \\
\hline HCCR\cite{Chiang1} & One ISP & $\star$ & $\star$ & $\surd$ & $\surd$ & $\varnothing$ \\
\hline HCCR\cite{Chiang1} & More ISPs & $\star$ & $\star$ & $\surd$ & $\star$ & $\varnothing$ \\
\hline \hline
\end{tabular}\\
\caption{Comparisons of Existing Work on Net non-neutrality}
\label{table:comp}
\end{table}


Game theoretic modeling of neutral or non-neutral  networks  may
consider as actors that are involved in strategic interactions
not only service providers and content providers but also the
users as well as the advertisement sector that often represents
a major source of revenue to the content providers.
Not all game theoretic models studied the net neutrality problem
from a non-cooperative point of view. When considering
the relation between service providers and content providers,
some researchers have considered cooperative mechanisms to
regulate the price that one provider pays to the other.
The Nash bargaining solution was considered in \cite{claudia,R9,networking2011}
where as the Shapley value was considered in \cite{shap1,shap2}.

The Nash bargaining paradigm is also known as the proportional
fair rate allocation in the traffic engineering \cite{kelly}. It is the unique way of transferring utilities
that satisfies a well known set of four axioms \cite{nash} related to fairness.
In \cite{shap1}-\cite{shap2} the Shapley value (which is known to have some
fairness properties \cite{e-winter}) has been used for deciding how
revenues from end users should be split between the service and the content
providers. Interestingly, it is the service provider that is seen to
be the one that has to pay the content provider, which reflects the
fact that the benefits of the service providers are obtained thanks
to the presence of the content provider (assuming that users subscribe
to the ISP in order to access the content of the CPs).

\subsection{Literature Survey}

We classify the existing work mainly based on types of game models that are used, i.e. noncooperative
and the cooperative games. For the models concentrated on noncooperative
price competition, we subdivide them into two classes. One of them assumes the same QoS
for the packets of all content providers. The other, on the contrary, allows
an ISP to provider premier QoS for the content providers that agree to pay to the ISP.
Beside the game oriented studies, we further describe a work that
considers the neutrality issue in network utility maximization.

\subsubsection{Noncooperative Game without Considering QoS Differentiation}

Economides and Tag \cite{Economides09} proposed a model of two-sided market in which the ISPs
play the role of platform, collecting revenues from both the end users and the non-competing content providers.
The quantities of content providers and end users are assumed to be normalized continuums.
In the monopoly market, the demand generated by end users is increasing with regard to the quantity of the CPs,
while decreasing along with the access price of the ISP. A CP has a positive
externality of revenue from advertisers, at the cost of payment to the ISP.
According to their analysis, the ISP and the users are better off, but the CPs' surplus
as well as the social surplus are worse without network neutrality.
In a duopoly market with two ISPs and multi-homing CPs, the quantity of
end users subscribing to one ISP depends on not only
the strategy of this ISP, but also that of its opponent. By using non-cooperative game tool,
the authors find that the total surplus as well as the surplus
of the CPs and the ISPs are better off at the equilibrium under the neutral regulation. This is opposite
to the monopoly ISP market.

Musacchio, Schwartz and Walrand \cite{Walrand09} investigate a two sided market where the CPs and the ISPs
invest jointly on the network infrastructure. Each ISP is a monopoly over its end users and the CPs
can be contacted by all the users. The total click
rate (or flow rate equivalently) to the CPs
is strictly increasing with regard to the investment of all the CPs and the ISPs. For a CP, a larger
investment will attract more clicks, hence bringing more revenues from the advertisers.
The major performance measure studied in this work is the social surplus.
The authors indicate that
the ratio between parameters characterizing advertising rates and end user price sensitivity plays
a key role in choosing the one-sided or the two-sided pricing. If this ratio is either low or high,
the two-sided pricing is more favorable, and vise versa.

In \cite{Zhang1}, Zhang et al. study the competition and innovation of the service-oriented Internet.
This service-oriented Internet can be regarded as a two sided market composed of two CPs and two ISPs.
The CPs charge end users based on their usage, while the ISPs charge them flat rate fees.
The CPs engage in a Cournot competition where the price is determined
by the total demand from the users. The ISPs engage in a Bertrand game so that they compete
over the side payment from the CPs (while not the end users). The Cournot and the Bertrand competitions
are tied together in a two-stage Stackelberg game. The authors
indicate that the update of an ISP
becomes profitable only when the increase of the marginal cost is upper bounded by an
appropriate gain in its market share.

Motivated by \cite{HW}, the authors of \cite{R8} investigate network non-neutrality with
a monopoly ISP, one CP and a number of end users.
The sources of income (other than side payments)
are payments of end users (to both the ISP and the CPs), and some third party payments (e.g.
publicity income) that the content providers receive. We formulate the price
competition as a noncooperative game. The CP's strategy is its charging to the users,
and the ISP's strategy is the charging from both the CP and the end user.
In \cite{R8} we find that if the ISP has the power to decide the
side payment, \underline{not only do the CP and the end users suffer, but also the} \underline{ISP's utility degrades.}
More precisely, we show that the only possible equilibrium would be
characterized by prices that will induce zero demand from the users.
This phenomenon does not occur if the CP's payment is fixed by some regulators, or
the ISP determines the payments from the CP and the users sequentially.

We extend \cite{R8} to incorporate the QoS of users provided by the ISP in the non-neutral model \cite{networking2011}.
This QoS measure is different from those in \cite{CBG,ChoiAndKim}
where more demands lead to a reduced QoS. We connect the QoS with
the incentive of ISP's investment. A larger demand from the users means a larger revenue, resulting in a
larger bandwidth provision of the ISP and a better QoS.
The authors introduce a parameter called \emph{relative price sensitivity}
to model the difference of demand sensitivities to the price of the ISP
and that of the CP.
When the price paid by the CP to the ISP for per-unit of traffic is a constant,
the qualitative impact of being non-neutral is decided
by the \emph{relative price sensitivity}. If this relative sensitivity is greater than 1,
the users value the service of the ISP more than that of the CP. We show that
a positive payment of the CP to the ISP leads to worse surplus of all parties involved
and worse QoS of end users. Our analysis reveals an implication that
the ISP may pay to the CP of high price sensitivity
so that the CP is able to reduce its service price. This type of reverse payment
is rarely discussed in the literature of network non-neutrality.

In \cite{R10}, we explore the effects of
content-specific (i.e.\ not \emph{application} neutral) pricing,
including multiple CPs providing different types of content.
Also, we consider competition among multiple providers of
the same type, including different models
consumer stickiness (inertia or loyalty).
In an on-going work, we are also considering providers'
infrastructure and operating costs (as in, e.g.\
 \cite{Walrand09}),  more complex models of end-user
demand and their collective social welfare, and  the
effects of  different  options for flat-rate pricing
 (e.g.\ \cite{Odlyzko01,ciss08}).

\subsubsection{Noncooperative Game with QoS Differentiation}

Hermalin and Katz \cite{HermalinAndKatz} consider the two sided market with a
monopolistic ISP and continuum of CPs and end users. They compare the levels of
profit, social welfare under neutral and non-neutral regime. In the non-neutral
regime they assume that the ISP can offer a range of differentiated
quality of connection qualities (e.g., an ISP can offer different combinations
of bandwidth, latency, and packet loss rate) to the CPs and charges them depending
on the type of quality of connections they opt for. Their analysis suggests that
any restriction on the ISP's choice to offer differentiated services will often
result in poor social welfare and it improves social welfare only under few
conditions. They further observe that the small scale content providers-the
ones who are intended to benefit from regulations- are almost always harmed
by the regulations. They extend the analysis of non-neutral regime to a Hotelling
duopoly ISP model and observe that welfare results of monopoly ISP carry over.

Economides and Hermalin \cite{EconomidesAndHermalin} study a two sided market
similar to the one in \cite{HermalinAndKatz} by considering the affect of network congestion.
They allow the amount of information purchased by the users to vary.
Their work shows restriction on granting or selling of priority services, i.e.,
neutral regime leads to superior social welfare. Further they show that the
incentive to invest is ambiguous under non neutral regime. The investment by the
ISP helps to improve the overall quality of the network  and thus reducing to some
degree the difference among the services offered at discriminated prices,
thus reducing the ISP's income.

Cheng et al. \cite{CBG} study a market with one ISP, two competing content providers and a finite
number of end users. The monopoly ISP provides two type of services,
the preferential and the non-preferential delivery. The content providers
can pay the ISP a fixed fee for preferential service, which implies a non-neutral network.
The authors model the QoS by the M/M/1 queueing delay, and the competition of the CPs
by a hotelling framework.
In this paper, the
principle with no regulation is beneficial to the surplus of the ISP,
while harmful to those of the CPs. The social benefit is improved
when one of the CP pays to the ISP, but remains unchanged when both CPs join the preferential service.
The non-neutrality might lead to better QoS for a majority of users and worse QoS of the others
if the social surplus is better.
The authors also observe that the ISP has less incentive for capacity expansion in
a non-neutral network.
This is because if the bandwidth is upgraded, more users experience less congestion and switch
to the non-preferential service. 
Thus, the difference of aggregate surplus of the ISP before and after capacity expansion
becomes smaller in a non-neutral regime
than in a neutral regime, resulting in less incentives to invest.




Choi and Kim \cite{ChoiAndKim} study the investment incentives with and without network regulation.
They consider a monopolist ISP and two CPs. Prioritization of delivery of packets by assigning
``fast lane'' to one of the CPs who agree to pay to the ISP is considered as the main mode of
the non neutrality. The congestion is taken into account by modeling the network as M/M/1 queue.
They study the  neutral and non neutral regime by comparing the market equilibrium in the short
run (fixed capacity) and investment incentives in long run. They observe that in the short run
the CPs will face the prisoner's dilemma to get access to the fast lane and will
be worse off. The social welfare improves in the non-neutral network when there is significant
asymmetry across the content providers. In the long run they argue that contrary to ISP's claim
that net neutrality regime will have adverse affect on their expansion, they may not have
investment incentive in the non-neutral regime. The CPs also may not have investment incentives
as they may fear that the ISP can expropriate some of the benefits made by them. Their analysis
yields ambiguous conclusions on investment incentives. 

Njoroge et al. \cite{asu} consider a network where two interconnected ISPs compete for
the users and the CPs over the quality and the price. The definition of
``neutrality'' is different from that of \cite{Economides09}. In a neutral model, a CP
pays to one ISP for Internet access, but does not pay to the other ISP without direct
connection. In a non-neutral model, a CP has to pay to the ISP without direct connection
in order to be reachable by its end users.
%
The inter-ISP link
is bandwidth limited so that the quality of a user-CP connection is decided by this bottleneck.
The authors model the price and quality competition as a six-stage sequential game and solve it
using backward induction. They show that the non-neutrality is able to improve the surplus of the ISPs, the CPs and the users.
The social surplus is also better, and the ISPs have higher incentives of investment.

\subsubsection{Regulation Mechanisms based on Cooperative Games}

Giving the full control of a market, an ISP can charge an arbitrary price from
CPs for the delivered contents.
However, there does not exist such a ``dictatorship'' status in reality.
Recent work introduces cooperative game tools, such as Nash bargaining game
and Shapley value, to study the revenue splitting issues among the players.

Shapley value is a well known concept in cooperative game theory
that provides a way of splitting revenues obtained by the cooperating players
\footnote{On the usefulness of this concept to split profits or
costs we can learn from \cite{binmore} where K Binmore writes \cite{binmore}:  ``I was once
summoned urgently to London to explain what the French government was talking
about when it suggested that the costs of a proposed tunnel under the English
Channel be allocated to countries in the European Union using the Shapley value''
}. It satisfies important properties like fairness, efficiency, symmetry, additivity, etc.
In \cite{shap1} Ma et al. explore Shapley pricing mechanism \cite{LloydShapley}
to share the revenues from Internet subscribers
among the service providers that peer each others' traffic.
The authors show that if Shapley value based revenue sharing is enforced at the global
level the selfish ISPs, at the Nash equilibrium point, will opt for strategies that
will maximize the aggregate network profits. Further extending their work, in \cite{shap2}
the authors consider three type of ISPs: content ISPs, eye ball ISPs and transit ISPs.
They obtain closed form expressions of the ISPs' revenues for the bipartite topologies (each
type of ISP nodes can be separated) and give dynamic programming procedures to
calculate the Shapley revenues for the general internet topologies.
With the Shapley value solution they suggest the appropriate pricing structure for
the differentiated services (non-neutral regime) that improves social welfare.

Saavedra in \cite{claudia} initially uses Nash bargaining game to study joint investment in
a non-neutral regime with one CP and two ISPs. The CP is able to
negotiate with one or the both ISPs via contracts. The author highlights
the impact of the CP's bargaining power on QoS agreements. If the
ISPs allow lower quality of services in the non-neutral regime,
the CPs with low bargaining power can enter into the exclusive contract
with ISP and improve their bargaining position.
Building upon \cite{claudia}, Altman et al. focus In \cite{R9}
on mechanisms of setting the price that one provider would pay
to another based on the Nash bargaining paradigm.
The authors address a problem and a model that deffer however, from
\cite{claudia}. Two bargaining concepts, the \emph{pre-bargaining} and the \emph{post-bargaining}
are presented to characterize the exact sequence of decisions that determine
the side payment as well as the other prices.
The first concept is used when
bargaining over the side payment takes place before the user prices
are determined.
The second one models the occurrence of bargaining after the user
price competition.
The authors point out under what situations the \emph{pre-bargaining} or the \emph{post-bargaining}
is preferable. They further study another aspect of
non-neutral behavior which is  the possibility for an ISP and a CP
to collude together which could result in better performances
for them but a worse performance for other competing CPs.

\subsubsection{Network Neutrality in Network Utility Maximization Issues}

In this survey, we do not restrict to the papers that use game theoretic models.
We mention one more reference whose model includes some aspects of competition.
Hande et al. \cite{Chiang1} consider the two-sided market in a quite different context. Instead of using
the predefined price-demand curve, they look into the network utility maximization
with the participation of the CPs. This model mitigates the key drawback of economic studies
that does not deal with the engineering aspect of rate allocation.
They define the utility of users as a function of flow rates. The concept ``non-neutrality''
refers to the restriction on the maximum price that the ISPs can extract from the CPs.
This price restriction exhibits different impacts on the profits concerning the power of the ISPs.
In a market of competitive ISPs, the users' benefit and the social benefit increase as the price
restriction is relaxed. If we understand neutrality as zero payment for per-unit of
delivered content, the non-neutral regime is always favorable in terms of the social profit and the users' profits.

\section{Conclusion}

We summarized some recent aspects in the development of the
Internet as well as in the debate on its neutrality.
We cover both   legal aspects, economic ones as well
as technological issues. We then present an overview of research papers on
the net non-neutrality problem. We focus in particular  on issues related
QoS.  As a universal service, the Internet is required to
guarantee certain QoS levels. We related this to
the legislation over network neutrality and to the question
of the incentives for investing in the infrastructure.
We then provided an overview of the research studies on network
neutrality issues with an emphasis on using game theoretical
tools.\\

\noindent \textbf{Acknowledgement}\\
This work was performed within the Web Science Institute of Avignon
University (SFR. ST des CSN). It was partly supported by the INRIA ARC Meneur
on Network Neutrality.



\begin{thebibliography}{99}

\bibitem{universal}
{\url{http://ec.europa.eu/information_society/policy/ecomm/current/consumer_rights/universal_service/index_en.htm}
}
\bibitem{arcep-may2010e}
ARCEP, ``Neutralit\'{e} des r\'{e}seaux: Les actes du colloque'', 2010.
{\url{http://www.arcep.fr/index.php?id=8652}}

\bibitem{comnet}
Sulan Wong, Eitan Altman, and Julio Rojas-Mora,
``Internet access: Where law, economy, culture and technology meet'',
\emph{Computer Networks}, Vol.55, No.2. 2011.

\bibitem{taxgoogle}
Jesse Drucker,
``Google 2.4\% Rate Shows How \$ 60 Billion Lost to Tax Loopholes'',
Oct 21, 2010, published in Bloomberg, available at \\
{\url{http://www.bloomberg.com/news/2010-10-21/google-2-4-rate-shows-how-60-}}
{\url{billion-u-s-revenue-lost-to-tax-loopholes.html}}


\bibitem{FrenchConsRes2010}
{Minist\`ere de l'\'Economie, de l'industrie et de l'emploi}.
\newblock {Consultation publique sur la neutralit\'e du Net}.
\newblock telecom.gouv.fr, June 21 2010.


\bibitem{InternetEU2010}
{Internet World Stats}.
\newblock Internet users in the european union, June 30 2010.
{\url{www.internetworldstats.com/stats9.htm}}.


\bibitem{Canada_Consultation:2009}
{Canadian Radio-television and Telecommunications Commission}.
\newblock {Telecom Regulatory Policy CRTC 2009-657}, 2009.


\bibitem{COMCAST_v_FCC:2010}
{United States Court of Appeals, District of Columbia Circuit}.
\newblock {COMCAST Corporation (Petitioner) v. Federal Communications
  Commission and USA (Respondents) and NBC Universal,
  \emph{et al.} (Intervernors), 600 F.3d 642}, 2010.


\bibitem{ZelnikToubonCerutti2010}
P.~Zelnik, J.~Toubon, and G.~Cerutti.
\newblock {Creation et Internet}.
\newblock Technical report, Ministre de la Culture et de la Communication,
  R\'epublique Fran\c{c}aise, 2010.


\bibitem{Google2010}
C\'ecile Ducourtieux.
\newblock {Le patron de Google re\c{c}u par Nicolas Sarkozy}.
\newblock {Le Monde.fr}, September 9, 2010.

\bibitem{FCC:2010c}
Federal Communications Commission.
\newblock Report and Order (FCC 10-201), 2010

\bibitem{EuroConsultNNRep2010}
European Commission: Information Society and Media Directorate-General.
\newblock Report on the public consultation on the open Internet and net neutrality in Europe, 2010.

\bibitem{France3061}
Jean-Marc Ayrault \emph{et al.}
\newblock Proposition de loi N° 3061 relative \`a la neutralit\'e de l'internet, 2010.

\bibitem{Kroes:2010b}
Neelie Kroes.
\newblock Who pays what? Broadband for all and the future of Universal Service Obligations, 2010.

\bibitem{Directiva136:2009a}
{European Parliament} and {Council of the European Union}.
\newblock {Directives 2009/136/EC, 2002/22/EC, 2002/58/EC of the European Parliament and of the Council. }

\bibitem{COM_2010_245}
{European Commission}
\newblock{A Digital Agenda for Europe} (COM(2010) 245), 2010.

\bibitem{EconSost2011}
{Ministerio de Econom\'ia y Hacienda del Reino de Espa\~na}
\newblock{Proyecto de Ley de Econom\'ia Sostenible}, 2011.

\bibitem{MinistryFinland:2009}
{Ministry of Transport and Communications of Finland}
\newblock{Decree on the minimun rate of a funcional Internet access as a universal service (732/2009)}, 2009.

\bibitem{Commission-of-the-European-Communities:2009}{European Commission}
\newblock {Telecom Reform 2009: Commission Declaration on Net Neutrality}

\bibitem{R7}
Sulan Wong, Julio Rojas-Mora and Eitan Altman,
``Public Consultations on Net Neutrality 2010'',
\emph{Proc. of NetCoop'10}, Ghent, Belguim, November 2010.



\bibitem{R8}
E.~Altman, P.~Bernhard, S.~Caron, G.~Kesidis,
J.~Rojas-Mora and S.L.~Wong.
``A Study of Non-Neutral Networks with Usage-based Prices'',
\emph{The 3rd ETM Workshop}. Amsterdam, 2010. Longer
version: INRIA research report 00481702.

\bibitem{R9}
E.~Altman, M.K.~Hanawal and R.~Sundaresan,
``non-neutral network and the role of bargaining power in side payments'',
NetCoop, Ghent, Belgium, Nov 2010.

\bibitem{networking2011}
E.~Altman, A.~Legout and Y.D.~Xu,
``Network Non-neutrality Debate: An Economic Analysis'',
to appear in IFIP Networking 2011.

\bibitem{R10} Application neutrality and a paradox of side payments, Stephane Caron, George Kesidis and Eitan Altman, The 3rd Int. Workshop on Re-Architecting the Internet (ReArch 2010), Nov. 30, Philadelphia, USA, collocated with ACM CoNEXT.

\bibitem{e-winter}
Eyal Winter, ``The Shapley value,'' Chapter 53 in {\em The Handbook of Game Theory, Vol. 3}. R.J.Aumann and S.Hart, North-Holland, 2002.

\bibitem{HW}
Robert Hahn and Scott Wallsten,
``The Economics of Net Neutrality,''
The Berkeley Economic Press Economists' Voice 3, 6, (2006), 1-7

\bibitem{shap1}
T.B.~Ma, D.M.~Chiu, J.C.S.~Lui, V.~Misra, D.~Rubenstein,
``Interconnecting eyeballs to content: A shapley value perspective on ISP
peering and settlement,''
\emph{Proc. of ACM NetEcon'08} , pp. 61-66,  2008

\bibitem{shap2}
T.B.~Ma, D.M.~Chiu, J.C.S.~Lui, V.~Misra, D.~Rubenstein,
``On cooperative settlement between content, transit
and eyeball internet service providers,''
\emph{Proc. of ACM CoNext'08}, New York, USA, 2008.

\bibitem{nash}
J.F. Nash, ``The bargaining problem,''
{\em Econometrica}, vol 18, pages 155--162, 1950.

\bibitem{kelly}
F.P. Kelly,  A. Maulloo and D. Tan,
``Rate control in communication networks: shadow prices,
proportional fairness and stability,''
{\em J. Oper. Res. Society}, Vol. 49, pp. 237--252, 1998.


\bibitem{claudia}
C.~Saavedra,
``Bargaining, power and the net neutrality problem,''
manuscript, presented at
NEREC Research Conference on Electronic Communications,
Edcole Polytechnique, 11-12 September 2009.


\bibitem{Walrand09}
J. Musacchio, G. Schwartz and J. Walrand,
``A two-sided market analysis of provider investment incentives
with an application to the net-neutrality issue'',
{\em Review of Network Economics}, {\bf 8}(1), 2009.

\bibitem{Odlyzko01}
A. Odlyzko.
``Internet pricing and history of communications''.
{\em Computer Networks}, {\bf 36}(5-6): pp. 493-518, Aug. 2001.

\bibitem{asu}
P. Njoroge, A. Ozdagler, N. Stier-Moses, G. Weintraub,
``Investment in two-sided markets and the net-neutrality debate,''
Decision, Risk, and Operations Working Papers Series, DRO-2010-05, Columbia Business School, July 2010.

\bibitem{ciss08}
G. Kesidis, A. Das, G. de Veciana,
``On Flat-Rate and Usage-based Pricing for Tiered Commodity
Internet Services'',
in \emph{Proc. CISS}, Princeton, March 2008.


\bibitem{economides}
Nicolas Economides,
``Net Neutrality, Non-Discrimination
and Digital Distribution of Content
Through the Internet'',
Journal of Law and Policy for the Information Society,
vol. 4, no. 2, pp. 209-233 (2008).

\bibitem{CBG}
Hsing K. Cheng, Subhajyoti Bandyopadyay and Hong Guo,
``The debate on Net Neutrality: A policy Perspective'',
Information Systems Research, March 1, 2010

\bibitem{Economides09}
N.~Economides and J.~Tag, ``Net Neutrality on the Internet: A Two-Sided Market Analysis''. NET Institute Working Paper No. 07-45;
Available at SSRN: http://ssrn.com/abstract=1019121

\bibitem{Chiang1}
P.~Hande, M.~Chiang, R.~Calderbank and S.~Rangan, ``Network Pricing and Rate Allocation with Content Provider Participation'',
\emph{Proc. of IEEE Infocom 2009}, pages: 990-998.

\bibitem{Zhang1}
Z.L.~Zhang, P.~Nabipay and A.~Odlyzko, ``Interaction, Competition and Innovation in a
Service-Oriented Internet: An Economic Model'', \emph{Proc. of IEEE Infocom 2010}.

\bibitem{HermalinAndKatz}
B.E.~Hermalin, M.L.~Katz, ``The Economics of product-line restrcitions with an applications to the neutrality debate'', AEI-brookings joint center for regulatory studies. Available at \url{http://ssrn.com/abstract=1003391}

\bibitem{EconomidesAndHermalin}
N.~Economides, B.E.~Hermalin, ``The Economics of Netwotk Neutrality'', NET Institute Working Papers, No.10-25, available at \url{http://works.bepress.com/economides/38/}

\bibitem{ChoiAndKim}
J.P.~Choi and B.C.~Kim, ``Net neutrality and Investment incentives'', to appear in Rand Journal of Economics.

\bibitem{JL}
K.~Jan and W.~Lukas, ``Network Neutrality and Congestion-Sensitive Content Providers: Implications for Service Innovation,
Broadband Investment and Regulation'', MPRA Paper No. 22095, 2010.

\bibitem{LloydShapley}
Lloyd S. Shapley, ``A Value for n-person Games". In Contributions to the Theory of Games, volume
II, by H.W. Kuhn and A.W. Tucker, editors.
Annals of Mathematical Studies v. 28, pp. 307¨C317. Princeton University Press, 1953.

\bibitem{binmore}
Ken Binmorme,
{\it Game theory, a very short introduction},
Oxford Univ Press, 2007

\end{thebibliography}
\end{document}